\documentclass[aps,prd,amssymb,preprintnumbers,notitlepage,nofootinbib]{revtex4-1}
\usepackage{graphicx}
\usepackage{bm}

\newcommand{\tfrac}{\frac}
\def\be{\begin{equation}}
\def\ee{\end{equation}}
\def\ba{\begin{eqnarray}}
\def\ea{\end{eqnarray}}
\def\bs{\begin{subequations}}
\def\es{\end{subequations}}

\usepackage{color}
\usepackage{bm}

\newcommand{\GB}{\mathcal{G}}
\newcommand{\de}{\mathrm{d}}

\newcommand{\B}{\nu}

\begin{document}

\title{Inevitable ghost and the degrees of freedom in $f(R,{\cal G})$ gravity}

\author{Antonio \surname{De Felice}}
\affiliation{Department of Physics, Faculty of Science, Tokyo University of Science, 1-3, Kagurazaka, Shinjuku-ku, Tokyo 162-8601, Japan}

\author{Takahiro \surname{Tanaka}}
\affiliation{Yukawa Institute for Theoretical Physics, Kyoto University, Kyoto 606-8502, Japan}

\begin{abstract}
The study of linear perturbation theory for general functions of the Ricci and Gauss-Bonnet scalars is done over an empty anisotropic universe, i.e.\ the Kasner-type background, in order to show that an anisotropic background in general has ghost degrees of freedom, which are absent on Friedmann-Lema\^itre-Robertson-Walker (FLRW) backgrounds. The study of the scalar perturbation reveals that on this background the number of independent propagating degrees of freedom is four and reduces to three on FLRW backgrounds, as one mode becomes highly massive to decouple from the physical spectrum. When this mode remains physical, there is inevitably a ghost mode. 
\end{abstract}

\date{\today}
\preprint{YITP-10-49, YITP-T-10-01}

\maketitle

\section{Introduction}
Modifications of gravity have been used in many contexts~\cite{review}. Perhaps the most famous is the first model of inflation introduced by Starobinsky in 1980~\cite{Starob}. However, in the last few years, modifications of gravity have been attracting attention as an alternative to quintessence~\cite{fRori}, in order to explain the late-time acceleration at large cosmological scales~\cite{fRviable,fRmatter,GaussBN}. The most general modifications of gravity whose Lagrangian is built using only the metric tensors so as not to introduce any extra vector or spin-2 degrees of freedom other than the graviton are such that the Lagrangian is given by 
\begin{equation}
  \label{eq:LG1}
  S=\frac{M_P^2}{16\pi}\int \de^4x\sqrt{-g}\, f(R,\GB)\, , 
\end{equation}
where $R$ and $\GB\equiv
R^2-4R_{\mu\nu}R^{\mu\nu}+R_{\mu\nu\alpha\beta}R^{\mu\nu\alpha\beta}$ are the Ricci ($R$) and Gauss-Bonnet ($\GB$) scalars, respectively. We have only two non-zero Lovelock scalars in four dimensions~\cite{PQR}. There are several recent studies on this type of models~\cite{DeFelice09,otherFRG}.

In Ref~\cite{DeFelice09}
the linear cosmological perturbation theory 
of these general modifications on a 
Friedmann-Lema\^itre-Robertson-Walker (FLRW) background was studied,
and the result
is that only one propagating mode for the scalar-type perturbations 
is present for a general
function $f$. 
Furthermore, such a mode has a scale-dependent
speed of propagation, depending on the wave number $k$ as well as 
the time-dependence through the background quantities. 
Furthermore, the 
background can be chosen so that this scalar mode is not a ghost. 
Each term in the Lagrangian of this scalar mode has 
only two temporal differentiations, 
whereas the square of the Laplacian 
operator appears, leading to a dispersion relation such as 
$\omega^2=B \, k^4/a^4$, where $B$ is
a function of the background quantities.
Even though the Lagrangian looks exotic, there is no ghost 
as long as the signature of the second 
temporal derivative term is
normal~\cite{ghost1}.

However, a few questions naturally arise by considering 
this result carefully. 
For example, in~\cite{DeFelice09} the studied background was
special, i.e.\ spatially homogeneous and isotropic. 
If we consider other inhomogeneous/anisotropic backgrounds, 
the term proportional to $k^4$, i.e.\ the operator $\triangle^2$, may 
shift to the term proportional to $\omega^4$, i.e.\ the operator
$\partial_t^4$, 
which would imply the existence of ghost degrees of
freedom. Therefore it is interesting to study the behavior of
this same theory on one of the simplest anisotropic backgrounds, the
Kasner-type solution.

There is another point which is tightly connected to the previous
one. The action (\ref{eq:LG1}) can be rewritten as 
\begin{equation}
S=\frac{M_P^2}{16\pi}\int \de^4x\,\sqrt{-g}\,\bigl[ F\,R+\xi\,\GB-V(F,\xi) \bigr], \label{action1}
\end{equation}
where $F$, and $\xi$ are auxiliary fields. 
By integrating out these auxiliary fields, 
we can verify that the action (\ref{action1}) is equivalent to
(\ref{eq:LG1})~\cite{Hanlon}.  
On the FLRW background 
among the two extra scalar fields introduced in Eq.~(\ref{action1}), 
one linear combination of their perturbations does not have kinetic term. 
Here, it is totally unclear if such a feature is common
to all backgrounds or it only happens on the FLRW one. 
In fact, as we shall see later, 
both modes propagate independently on Kasner-type backgrounds. 
In this paper, we will also see that having one more propagating
field implies that the $\omega^2\propto k^4$ dispersion relation 
returns back to a normal one, i.e.\ $\omega^2\propto k^2$. 
But how can we explain this apparent reduction of the propagating degrees of freedom from one background to another? As we shall see later, the determinant of the kinetic operator for the perturbation modes reduces to zero as we take the limit of the FLRW backgrounds. At the same time, the mass for the mode whose kinetic term vanishes on the FLRW backgrounds in general blows up to infinity. Therefore, in the FLRW limit this extra mode decouples from the physical spectrum to be integrated out, and the number of effective propagating modes is reduced by one.


\section{Background models}
\subsection{Kasner-type backgrounds}

For the reasons mentioned above, we study a simplified Kasner-type anisotropic
background, which can be written as
\begin{equation}
  \label{eq:Kmet}
  \de s^2=-\de t^2+f_1(t)\, \de x^2+f_2(t)\,(\de y^2+\de z^2)\, .
\end{equation}
We analyze the behavior of the perturbations at linear order, by
using $1+1+2$ decomposition according to the symmetry of the 
background metric. 
From the symmetry in $y-z$ plane, in general, 
the perturbation can be decomposed into three even-parity modes and one odd-parity mode. 
One of the three even-parity modes has a vanishing kinetic term 
in the FLRW limit, where the mode decouples since its mass becomes 
infinitely large. 

The equations of motion for the system
(\ref{action1}) are 
\begin{eqnarray}
  \frac{\partial V}{\partial F}&=&R\,,\qquad
  \frac{\partial V}{\partial \xi}=\GB\,,\\
&&F\,G_{\mu\nu}=\Sigma_{\mu \nu}\, ,
\end{eqnarray}
where $\Sigma_{\mu \nu}$ is the tensor defined by
\begin{eqnarray} 
&\Sigma_{\mu \nu}&=\nabla_\mu \nabla_\nu F-g_{\mu \nu} \Box F+2R \nabla_\mu \nabla_\nu \xi-2g_{\mu \nu} R \Box \xi-4R_\mu{}^{\lambda} \nabla_\lambda \nabla_\nu \xi \nonumber \\
&&\quad -4R_\nu{}^{\lambda} \nabla_\lambda \nabla_\mu \xi 
+4R_{\mu \nu} \Box \xi+4 g_{\mu \nu} R^{\alpha \beta} \nabla_\alpha \nabla_\beta \xi+4R_{\mu \alpha \beta \nu} \nabla^\alpha \nabla^\beta \xi-\tfrac{1}{2}\,g_{\mu \nu} V \,. 
\label{enemom}
\end{eqnarray}
On the Kasner-type background we have
\begin{eqnarray}
  &R&={\frac {2\, \dot f_2  \dot f_1 f_1 f_2 +4\, \ddot f_2  f_1 ^{2}f_2
 +2\, \ddot f_1 f_1  f_2 ^{2}- \dot f_1 ^{2} f_2 ^{2}- \dot f_2 ^{2} f_1
 ^{2}}{2 f_2 ^{2} f_1 ^{2}}}\, , \\
&\GB&={\frac { \dot f_2  \left( 2\,f_2 f_1  \dot f_2 \ddot f_1 -2\,f_1
 \dot f_2 ^{2}\dot f_1 +4\,f_2 f_1  \dot f_1 \ddot f_2 -f_2  \dot f_2
 \dot f_1 ^{2} \right) }{2 f_1 ^{2} f_2 ^{3}}}\, .
\end{eqnarray}
We can use the ($tt$)-component of Einstein equations in order to write
$V(t)$ in terms of the other background quantities as
\begin{eqnarray}
  V(t)={\frac {2\,F  \dot f_2 f_2 \dot f_1 +6\, \dot\xi  \dot f_2
 ^{2}\dot f_1 +F  \dot f_2 ^{2}f_1 +4\, \dot F  \dot f_2 f_2 f_1 +2\,
 \dot F  \dot f_1  f_2 ^{2}}{2 f_2 ^{2}f_1 }}\, .
\end{eqnarray}
Furthermore, one can use the ($xx$)- and ($yy$)-components of 
Einstein equations in order to write $\ddot F$ and $\ddot\xi$ in terms
of the other background quantities as follows:
\begin{eqnarray}
\ddot F&=&\frac1{4f_1  f_2 ^{2} \,(  \dot f_2 f_1 -f_2 \dot f_1  )}
 \,[-2\, \dot f_2  f_1 ^{2}f_2  \ddot f_2 F +2\, \dot f_2 ^{2} f_1
 ^{2}f_2 \dot F -2\, \dot f_2 f_1 F  \ddot f_1  f_2 ^{2}-4\, \dot f_2
 ^{2} \dot\xi f_2  \dot f_1 ^{2}+4\, \dot f_2 ^{3}f_1  \dot\xi \dot f_1
 - \dot f_2  f_2 ^{2} \dot f_1 ^{2}F\nonumber\\
&&\quad - \dot f_2 ^{2}f_1  \dot f_1 f_2 F +4\, \dot f_2 f_1  \dot\xi f_2  \dot f_1 \ddot f_2 -4\,f_1 f_2 \dot\xi  \ddot f_1  \dot f_2 ^{2}-2\, \dot F  \dot f_1 ^{2} f_2 ^{3}+ 2\, \dot f_2 ^{3} f_1 ^{2}F +4\, f_1  f_2 ^ {2} \dot f_1  \ddot f_2 F ]\,, \\
\ddot\xi&=&-\frac1{4 \dot f_2 f_1 f_2 \,(  \dot f_2 f_1 -f_2 \dot f_1
 )}\,[2\,F  \ddot f_2  f_2 ^{2} f_1 ^{2}+2\, \dot F  \dot f_2  f_2 ^{2}
 f_1 ^{2}- 2\,f_1 F  \ddot f_1  f_2 ^{3}+2\, \dot f_1 ^{2} \dot\xi  \dot
 f_2 f_2 ^{2}+2\,f_1  \dot\xi  \dot f_1  \dot f_2 ^{2}f_2 + \dot f_1
 ^{2}F  f_2 ^{3}\nonumber\\
&&\quad +8\, f_1 ^{2} \dot\xi  \ddot f_2  \dot f_2 f_2 -4\, f_1 ^{2}
 \dot\xi  \dot f_2 ^{3}-2\,f_1  \dot F  \dot f_1  f_2 ^{3}-F  \dot f_2
 f_2 ^{2}f_1 \dot f_1 -4\,f_1  f_2 ^{2} \dot\xi  \ddot f_2 \dot f_1
 -4\,f_1  f_2 ^{2} \dot\xi  \ddot f_1 \dot f_2 ]  \, .
\end{eqnarray}

\subsection{Almost FLRW backgrounds}

For practical construction of cosmological models, 
we are particularly interested in the behavior
of these theories on the backgrounds close to FLRW. 
Therefore it is sometimes convenient to consider Kasner-type 
backgrounds which are described by homogeneously perturbed FLRW as 
\begin{eqnarray}
  f_1&=&a(t)^2+\delta f_1(t)\,,\\
  f_2&=&a(t)^2+\delta f_2(t)\,,\\
  F&=&F_0(t)+\delta F(t)\,,\\
  \xi&=&\xi_0(t)+\delta \xi(t)\, .
\end{eqnarray}
At the linear order in this perturbation, 
combining the ($xx$)- and ($yy$)-components of the Einstein
equations leads to a closed differential 
equation for the field
\begin{equation}
  \label{eq:delf1f2}
  \delta\equiv\delta f_1-\delta f_2\,,
\end{equation}
which reads
\begin{equation}
  \label{eq:dddlll}
  \ddot\delta+2\B\,\dot\delta+M^2\,\delta=0\, ,
\end{equation}
where
\begin{eqnarray}
  \B(t)&=&-\tfrac12H+\frac1{2(F_0+4H\dot\xi_0)}\,\frac{\de}{\de t}(F_0+4H\dot\xi_0)\,,\\
M^2(t)&=&-2\dot H-2H^2-\frac {2H}{F_0+4H\dot\xi_0}\,\frac{\de}{\de t}(F_0+4H\dot\xi_0)\, ,
\end{eqnarray}
with $H:=\dot a/a$. 
The differential equation admits the general solution
\begin{equation}
\label{eq:soldd}
\delta=c_1\,a(t)^2\int^t \frac{{\rm d}t'}{a(t')^3\,[F_0(t')+4H\dot\xi_0(t')]}+c_2\, a(t)^2\, .
\end{equation}
The second solution represents just the constant shift of the scale of spatial coordinates. Namely it is just a gauge mode. The first physical solution decays, compared to the $a^2$ term, when the term $a^3(F_0+4H\dot\xi_0)$ increases in time faster than $t^1$. In this case $f_1$ gets closer to $f_2$, and the Kasner-type background reduces to FLRW.

\section{Perturbation Analysis}
\subsection{Reduction of the action for the even-parity modes}
We begin with the even-parity modes, which here mean the scalar-type
perturbations in the two-dimensional space spun by $y-z$ plane. In order
to achieve this goal we consider the perturbation
given by ~\cite{cosmoreview}
\begin{eqnarray}
  F&=&F(t)+\delta F(t,\bm{x})\, ,\\
  \xi&=&\xi(t)+\delta\xi(t,\bm{x})
  \, ,\\
  \de s^2&=&-[1+\alpha(t,\bm{x})]\,\de t^2+\partial_x\chi(t,\bm{x})\,\de t\,\de
 x+\partial_y\zeta\,\de t\,\de y+\partial_z\zeta\,\de t\,\de
 z+[f_1+\beta(t,\bm{x})]\,\de x^2+f_2(t)\,(\de y^2+\de z^2)\, .
\end{eqnarray}
We have fixed the gauge so that the perturbations of the spatial metric in the two dimensional subspace spun by $y$ and $z$ vanish. 
We will find it convenient to make a field redefinition, 
\begin{equation}
\psi=\delta F+\frac{\dot f_1^2}{f_1^2}\,\delta\xi\, ,
\end{equation}
Expanding the action to the second order, and introducing a Fourier
expansion for the fields with the wavenumber $\bm{k}$, we
eliminate the perturbations of lapse and shift ($\alpha$, $\chi$ and
$\zeta$) by using the constraint equations obtained by the variations of
the action with respect to these non-dynamical variables. Then, one
reaches the result
\begin{eqnarray}
  S=\tfrac12\,\int \de t\, \de^3
 k(f_1f_2^2)^{1/2}\left[A_{ij}\dot\phi_i\dot\phi_j-B_{ij}\dot\phi_i\phi_j
 -q^2\,C_{ij}\phi_i\phi_j-D_{ij}\phi_i\phi_j\right].
\end{eqnarray}
where $\phi_i=(\beta,\psi,\delta\xi)$ and we have introduced $q\equiv
(k_y^2+k_z^2)^{1/2}$. All the matrices $A, B, C$ and $D$ depend on the momentum only through the ratio $\gamma\equiv k_x/q$. Only the matrix $B$ is antisymmetric, while the others are all symmetric.  

\subsection{Degrees of freedom and the signature of kinetic term}

It is important to establish the sign of the eigenvalues of the kinetic matrix $A$. If they are not all positive, it means that there exist ghost-like degrees of freedom. The first key issue is that the matrix $A$ is not degenerate on general backgrounds. Namely, 
\begin{equation}
\det A=
-\frac{4\dot f_2^4\, \Delta_1}{f_1f_2^2\,J^2}
\left(\frac{\dot f_1}{f_1}-\frac{\dot f_2}{f_2}\right)^{\!2}\, , 
\label{eq:detA}
\end{equation}
does not vanish. Here we defined
\begin{eqnarray}
\Delta_1&=&f_1\,F+2\dot f_1\,\dot\xi \label{eq:defDelt1}\,, \\
\label{eq:defJ}
 J&=&2 \dot F \gamma^2 f_2^2+2 \dot F f_1 f_2+\dot f_1 F f_2+6 \gamma^2 \dot f_2^2 \dot\xi+2 F \dot f_2 \gamma^2 f_2+6 \dot f_1 \dot\xi \dot f_2+\dot f_2 F f_1\, .
\end{eqnarray}
 Thus, the number of independent degrees of freedom present in the even-parity perturbation is three. 

We notice that $\det A$ vanishes in the FLRW limit, $f_1\to f_2$. This implies a reduction of the number of the degrees of freedom in this limit. In fact, in this limit both $A_{13}$, and $A_{23}$ become of order $O(\delta)$; whereas $A_{33}$, like the determinant, is of order $O(\delta^2)$, as shown by the following expressions
 \begin{eqnarray}
 A_{13}&=&\frac {2Hc_1}{ a ^{5} \,( F +4\,H \dot\xi ) \left( {\gamma}^{2}+1 \right) ^{2} ( 2\,F H +\dot F +12\, \dot\xi H ^{2} ) ^{2}}\,  ( -40\, H ^{3}F \dot\xi -4\,H F  \dot F {\gamma}^{2}-48\, H ^{4}{\gamma}^{2} \dot\xi ^{2}-144\, H ^{4} \dot\xi ^{2}\nonumber\\
&&-8\, H ^{2} \dot F {\gamma}^{2}\dot\xi -6\, H F \dot F + \dot F ^{2}{\gamma}^{2}-2\, H ^{2} F ^{2}- \dot F ^{2}-16\, H ^{3}F {\gamma}^{2}\dot\xi -32 \, \dot F  H ^{2}\dot\xi  )\,,\\
A_{23}&=&-{\frac { 4c_1\,H ^{2} \,( -8\,F {\gamma}^{2}H -24\, \dot\xi {\gamma}^{2} H ^{2}+2\,{\gamma}^{2}\dot F -36\, \dot\xi H ^{2}-8\,F H -\dot F  )}{( 2\,F H +\dot F +12\, \dot\xi  H ^{2} ) ^{2} ( {\gamma}^{2}+1 )  \,( F +4\,H \dot\xi )  \,a ^{3}}}\,,\\
 A_{33}&=&-\frac {16c_1^2 H ^{2}}{a ^{6} ( F +4 \,H \dot\xi  ) ^{3} ( {\gamma}^{2}+1 ) ^{2} ( 2\,F H +\dot F +12\, \dot\xi  H ^{2}) ^{2}} \,(  \dot F ^{2}{\gamma}^{2}-48\, H ^{3}F {\gamma}^{2}\dot\xi +4\,H F  \dot F {\gamma}^{2}-16\,F  H ^{3}{\gamma}^{4}\dot\xi\nonumber\\
&& +24\, H ^{2} \dot F {\gamma}^{2}\dot\xi -48\, H ^{4}{\gamma}^{2} \dot\xi ^{2}-4\, H ^{2}{\gamma}^{4} F ^{2}-H F \dot F -4\, \dot F H ^{2}\dot\xi -8\, F ^{2}{\gamma}^{2} H ^{2}-44\, H ^{3}F \dot\xi -5\, H ^{2} F ^{2}-96\, H ^{4} \dot\xi ^{2}\nonumber\\
&&+4\,H  \dot F F  {\gamma}^{4}+16\, H ^{2} \dot F {\gamma}^{4}\dot\xi)\, ,
 \end{eqnarray}
(recall that $c_1$ is the magnitude of $\delta=\delta f_1-\delta f_2$), whereas the determinant tends to
\begin{equation}
  \det A=-\frac{16 H^4 c_1^2}{a^{10} (F+4 H \dot\xi) (\gamma^2+1)^2 (2 F H+\dot F+12 \dot\xi H^2)^2}\, .
\end{equation}
The fact that the sign of $\det A$ is opposite to the one of $F+4H\dot\xi$ shows that on Kasner-type backgrounds close to FLRW models, at least one mode becomes a ghost as $F+4H\dot\xi> 0$ is required for the tensor modes not to be ghost in FLRW backgrounds~\cite{DeFelice09}.

The matrix $A$, in the FLRW limit ($c_1\to0$), becomes block diagonal, i.e.\ it is composed of a 2$\times$2 matrix and 1$\times$1 matrix. The latter is zero valued. The 2$\times$2 submatrix is non-degenerate and finite. The sign of its eigenvalues are read from 
 \begin{eqnarray}
 A_{22}&=&\frac{12 H^2 (F+4 H \dot\xi)}{(2 F H+\dot F+12 \dot\xi H^2)^2
  }\, ,\\
 A_{11}A_{22}-A_{12}^2&=&\frac{3 (F+4 H \dot\xi)^2 H^2 }{(\gamma^2+1)^2 (2 F H+\dot F+12 \dot\xi H^2)^2\, a^4}\, .
 \end{eqnarray}
Since both of these two factors are positive (assuming $F+4H\dot\xi
>0$), the (unique) ghost mode in this limit is identified with the third field, which is $\delta\xi$ in the present calculation. We note that the important point to obtain the property that the kinetic term of the third field vanishes as $\delta^2$ is in how we choose the first and the second fields $\beta$ and $\psi$. As we will see below, the ghost mode $\delta\xi$ can be integrated out from the action near the FLRW backgrounds.

\subsection{The Laplacian operator and the $q^4$ behavior on FLRW}

Let us restrict our attention to better understand the FLRW limit for
the Laplacian matrix $C$. In particular we are interested in the elements
regarding the coupling of $\delta\xi$ with itself and with the other two fields. We find
  \begin{equation}
    \label{eq:detCC}
    \det C=-\frac{64\, (F+4 \ddot\xi)\, (1+\gamma^2)\, \dot H^2 H^2}{(2 F H+\dot F+12 H^2 \dot\xi)^2\, a^{10}}\, ,
  \end{equation}
and
\begin{eqnarray}
  \label{eq:C13}
  C_{13}&=&-\frac{4 (\dot F+4 \dot\xi H^2) \dot H}{(12 \dot\xi H^2+2 F H+\dot F) \, a^4}\,,\\
  C_{23}&=&\frac{16(\gamma^2+1) \dot H H}{(12 \dot\xi H^2+2 F H+\dot F)\, a^2}\, ,\\
  C_{33}&=&\frac{64 c_1 \dot H (2 \gamma^2+1) H^2}{(12 \dot\xi H^2+2 F H+\dot F) (F+4 H \dot\xi)\, a^5}\, .
\end{eqnarray}
 Therefore, also the $q^2$-self-coupling vanishes on FLRW, but the
 $q^2$-couplings with the other fields do not. It is of crucial importance whether the remaining term quadratic in $\delta \xi$, the mass term $D_{33}$, vanishes or not. $D_{33}$ is given by
\begin{eqnarray}
  \label{eq:D33}
  D_{33}&=&\frac{(\dot F+4 \dot\xi H^2)^2}{\dot\xi^2}\,\frac{\partial^2V}{\partial F^2}
-\frac{6(\dot F+4 \dot\xi H^2) }{\dot\xi^2}\,(\ddot H+4H\dot H)
+\frac{48 H \dot H^2}{\dot\xi}
-\frac{768 H^4 \dot H^2 (F+4 H \dot\xi)}{(12 \dot\xi H^2+2 F H+\dot
F)^2}\, , 
\end{eqnarray}
which does not vanish in general\footnote{The expression $\left(\frac{\partial^2V}{\partial F^2}\right)_{\!\xi\xi}$ can be rewritten in terms of second derivatives of $f$ in action (1), as follows $\left(\frac{\partial^2V}{\partial F^2}\right)_{\!\xi\xi}=f_{\GB\GB}/(f_{RR}f_{\GB\GB}-f_{R\GB}^2)$, where $f_{RR}\equiv\left(\frac{\partial^2f}{\partial R^2}\right)_{\!\GB\GB}$ etc.}. This implies that in the Lagrangian in the FLRW limit the only term quadratic in $\delta\xi$ is its mass. It should be noticed that in constructing cosmological models we have to require $F\approx 1$ at early times. Then, we have $\dot F\ll H$ and $H\dot\xi\ll1$. This implies that $D_{33}/H^4$ in general becomes very large at early times. This fact may help integrating out the ghost, as it naturally gives a very large mass to $\delta\xi$ compared to its kinetic term. The part of the Lagrangian related to $\delta\xi$ is written as
\begin{equation}
{\cal L}\ni-\tfrac12\,{a^3}\,D_{33}\,\delta\xi^2-
a^3 \delta\xi\,\bigl(
q^2\,C_{13}\,\,\beta+B_{13}\dot \beta
+q^2\,C_{23}\,\psi+B_{23}\dot \psi
\bigr)+\dots+O(\delta)\, .
\end{equation}
Integrating out $\delta\xi$, with the aid of the relation $C_{13}B_{23}=C_{23}B_{13}$, we obtain the terms proportional to $q^4$, which were found in \cite{DeFelice09}.

\subsection{Odd-parity modes}

Let us consider the odd-parity modes, i.e.\ the modes that are composed of divergence-less vectors in terms of two dimensions spun by $y$ and $z$. Perturbations of all four-dimensional scalars are zero for odd-parity modes. The perturbation of odd-parity modes appears only in the metric perturbation as  
\begin{equation}
\label{eq:vect}
  v_i=h_{0i}\,,\qquad w_i=h_{xi}\,,\qquad h_{ij}=0\, 
\end{equation}
where $i,j=y$ or $z$, and $v_i$ and $w_i$ are divergence-less vector. We used one gauge degree of freedom belonging to the odd-parity modes to eliminate the perturbation in ($ij$)-components. 

As in the case of even-parity modes, we erase the shift perturbation $v_i$ by using the constraint equation obtained by taking the variation with respect to $v_i$ itself. Then, we obtain 
\begin{equation}
S=\tfrac12\int \de t\de^3 k \, f_1^{1/2}f_2\,
A_{(o)}\left[\dot w_i\dot w_i-c_{(o)}^2\,{k^2\over f_1}
 \,w_iw_i-m^2_{(o)}\,w_iw_i\right]\, ,
\end{equation}
where
\begin{eqnarray}
A_{(o)}&=&\frac{\Delta_1 \Delta_2}{f_1f_2^2
(\Delta_1+\gamma^2\Delta_2)
}\,,\label{eq:Ao}
\\
c_{(o)}^2&=&\frac{(1+\gamma^2)^{-1}
(\Delta_1+\gamma^2\Delta_2)
}{\Delta_2 \Delta_1 \dot f_2 (\dot f_2 f_1-f_2 \dot f_1) f_2} (-2 F \ddot f_2 f_2^2 f_1^2+F f_1^2 \dot f_2^2 f_2+2 F f_1 \ddot f_1 f_2^3-\dot f_1^2 F f_2^3-2 \dot F \dot f_2 f_2^2 f_1^2\nonumber\\
&&+4 f_1^2 \dot\xi \dot f_2^3-8 f_1^2 \dot\xi \ddot f_2 \dot f_2 f_2+4 f_1 f_2^2 \dot\xi \ddot f_2 \dot f_1-2 f_1 \dot\xi \dot f_1 \dot f_2^2 f_2+4 f_1 f_2^2 \dot\xi \ddot f_1 \dot f_2+2 f_1 \dot F \dot f_1 f_2^3-2 \dot f_1^2 \dot\xi \dot f_2 f_2^2)\,,\\
m_{(o)}^2&=&\tfrac12 \,(-8 f_2^2 F \dot f_2 f_1^2 \ddot f_1 \dot\xi \dot f_1+8 f_2 \dot f_2^2 f_1^2 \dot\xi^2 \ddot f_2 \dot 
f_1^2-4 F f_2 \dot f_2^2 f_1^3 \gamma^2 \dot\xi \ddot f_2-4 f_2^2 F \dot f_2^2 f_1^2 \gamma^2 \dot\xi \ddot f_1
+4 F^2 f_1^2 \gamma^2 \dot f_1 \ddot f_2 f_2^3\nonumber\\
&&-8 f_2^2 \dot f_2 f_1 \dot\xi^2 \ddot f_1 \dot f_1^2-2 F^2 f_2^2 \dot f_2 f_1^3 \gamma^2 \ddot f_2+2 F^2 \dot f_2^3 f_1^3 \gamma^2 f_2+4 F \dot f_2^4 f_1^3 \gamma^2 \dot\xi-2 F^2 f_2^4 f_1 \gamma^2 \dot f_1 \ddot f_1+8 f_2^2 F f_1^2 \dot\xi \ddot f_2 \dot f_1^2\nonumber\\
&&-8 f_2 \dot f_2^3 f_1^2 \gamma^2 \dot\xi^2 \ddot f_1-4 \dot f_2 f_1 \gamma^2 \dot\xi \dot f_1^2 \dot F f_2^3+8 f_2^2 \dot f_2^2 f_1^2 \gamma^2 \dot\xi \dot f_1 \dot F+4 F f_2^3 \dot f_2 f_1^2 \gamma^2 \dot f_1 \dot F-6 F f_2 \dot f_2^3 f_1^2 \gamma^2 \dot\xi \dot f_1\nonumber\\
&&+12 f_2^2 F \dot f_2 f_1^2 \dot\xi \ddot f_2 \dot f_1 \gamma^2-4 F f_2^3 \dot f_2 f_1 \gamma^2 \dot f_1 \dot\xi \ddot f_1-2 F f_1 \gamma^2 \dot f_1^2 \dot F f_2^4+2 F f_2^3 \dot f_2 \gamma^2 \dot\xi \dot f_1^3-2 F f_2^2 \dot f_2^2 f_1^3 \gamma^2 \dot F\nonumber\\
&&-4 f_2 \dot f_2^3 f_1^3 \gamma^2 \dot\xi \dot F-3 F^2 f_2^2 \dot f_2^2 f_1^2 \gamma^2 \dot f_1+8 f_1 f_2^2 \dot\xi^2 \ddot f_2 \dot f_1^3+2 F^2 f_2^2 f_1^3 \ddot f_2 \dot f_1-2 F^2 f_2^2 \dot f_2 f_1^3 \ddot f_1+F^2 f_2^4 \gamma^2 \dot f_1^3)\nonumber\\
&&\times[(F f_1+2 \dot f_1 \dot\xi) (F f_1+2 \dot\xi \gamma^2 \dot f_2+2 \dot f_1 \dot\xi+f_2 F \gamma^2) (\dot f_2 f_1-f_2 \dot f_1) f_1 f_2^2]^{-1}\,,
\end{eqnarray}
and $\Delta_2=f_2\,F+2\dot f_2\,\dot\xi$. Here, $c_{(o)}$ looks divergent in the FLRW limit, but in fact it is
not. In the FLRW limit we find
\begin{eqnarray}
  A_{(o)}&=&\frac{(F+4H\dot\xi)}{a^4 (1+\gamma^2)}\,\\
  c_{(o)}^2&=&\frac{F+4\ddot\xi}{F+4H\dot\xi}\, ,  \label{eq:gravTT}\\
  m^2_{(o)}&=&-\frac{2 (8 \dot H H \dot\xi+\dot H F+4 H^3 \dot\xi+H^2 F+H \dot F+4 H^2 \ddot\xi)}{F+4 H \dot\xi}\, .
\end{eqnarray}
and $c_V^2$ reduces here to the speed of the tensor modes in the FLRW background. 

\section{Implication of the perturbation analysis around Kasner-type backgrounds}
\subsection{Inevitable ghost}

 Here we show that the no-ghost condition $A_{(o)}>0$ with $A_{(o)}$ given in Eq.~(\ref{eq:Ao}) is incompatible with another no-ghost condition ${\det A}> 0$ for the even-parity modes. From Eq.~(\ref{eq:detA}), we find that $\Delta_1<0$ is required for the absence of ghost. Then, if $\Delta_2<0$, $A_{(o)}$ becomes negative  for any $\gamma$, and hence the odd-parity modes becomes ghost. Even if $\Delta_2>0$, $A_{(o)}$ becomes negative for a sufficiently large $\gamma$, and hence at least a part of odd-parity modes becomes ghost. Therefore we conclude that the presence of ghost is inevitable for generic models of $f(R,\GB)$ once we consider Kasner-type backgrounds. This is the main claim of this paper. 

However, as we have seen above, the perturbative action for such a mode does not completely vanish in general even when the kinetic term vanishes, e.g.\ the matrix $D$ is not degenerate even in the FLRW limit. Therefore this ghost mode does not becomes strongly coupled in the FLRW limit, but it becomes infinitely massive. Thus, such a mode is to be integrated out as long as backgrounds are chosen to be sufficiently close to the FLRW universe. In this sense, such models might be harmless in spite of the presence of ghost. 

\subsection{Particular cases with only three degrees of freedom}

We will consider here the case for which one has the condition
\begin{equation}
  \label{eq:delt}
  \frac{\partial^2 f}{\partial R^2}\,\frac{\partial^2 f}{\partial
  \GB^2}-\left(\frac{\partial^2 f}{\partial
	  R\partial\GB}\right)^{\!2}=\left\vert{\partial(F,\xi)\over\partial(R,\GB)}\right\vert=0\, .
\end{equation}
In this case perturbations of $F$ and $\xi$ are not independent. For generic backgrounds, this is the case if ${\cal L}=F(\phi)R+\xi(\phi)\GB-V(\phi)$.

Below, let us consider special cases in which $f(R,\GB)$ is a function in the form of $\tilde f(R+\lambda {\cal G})$ with $\lambda$ being a constant In this case, 
\begin{equation}
  \label{eq:dxdf}
  \delta\xi={\dot\xi\over \dot F}\delta F\, ,
\end{equation}
generally holds. Therefore one can eliminate $\delta\xi$ from the perturbation action so that the number of even-parity degrees of freedom is reduced to two. Then, the no-ghost conditions for even-parity modes reduce to only two conditions 
\begin{eqnarray}
  \det A&=&-\frac{(2 \dot f_2+\dot f_1)^2}{\dot F^2 f_1^2 f_2^4 J^2}\,(2 \dot f_1 \dot\xi+f_1 F) (\dot F f_2^2+\dot\xi \dot f_2^2)\nonumber\\
  &&\times  (2 f_2^2 \dot F \dot f_1 \dot\xi-3 f_2^2 F \dot F f_1-8 f_2
 \dot f_2 \dot\xi f_1 \dot F-4 f_2 \dot f_2 \dot\xi F \dot f_1-6 \dot
 f_2^2 \dot f_1 \dot\xi^2+\dot f_2^2 F f_1 \dot\xi)> 0,\\
A_{22}&=&-\frac4{f_2^2 \dot F^2 f_1 J^2 (F f_2+2 \dot\xi\dot f_2)}(-f_2^2 \dot f_1^3 \dot\xi^2 F^2 \dot f_2^3+\dot\xi^2 f_1^2 \gamma^2 \dot f_2^6 F^2-12 \dot f_1 \dot\xi^4 \gamma^4 \dot f_2^7-24 \dot f_1^2 \dot\xi^4 \gamma^2 \dot f_2^6-2 f_2^3 \dot f_1^2 \dot\xi \dot F f_1 F^2 \dot f_2^2\nonumber\\
&&-6 f_2 \dot f_1 \dot\xi^2 \dot F f_1^2 F \dot f_2^4-f_2^3 \dot f_2^2 f_1^3 \dot F^2 F^2-12 f_2^2 \dot f_1^2 \dot\xi^2 \dot F f_1 F \dot f_2^3-3 f_2^2 \dot f_1 \dot\xi \dot F f_1^2 F^2 \dot f_2^3-4 f_2^3 \dot f_1 \dot F^2 f_1^2 F \dot\xi \dot f_2^2-f_2^5 \dot f_1^2 \dot F^2 f_1 F^2\nonumber\\
&&-f_2^4 \dot f_1 \dot f_2 f_1^2 \dot F^2 F^2-8 f_2^3 \dot f_1^2 \dot F^2 f_1 \dot\xi^2 \dot f_2^2-6 f_2^3 \dot f_1^3 \dot\xi^2 \dot F F \dot f_2^2-2 f_2 \dot f_1^2 \dot\xi^2 f_1 F^2 \dot f_2^4-16 f_2 \dot f_1^2 \dot\xi^3 \dot F f_1 \dot f_2^4+2 \dot\xi^3 f_1 \gamma^4 F \dot f_2^7\nonumber\\
&&-6 f_2^4 \dot f_1^2 \dot F^2 f_1 F \dot\xi \dot f_2-8 f_2 \dot\xi^3 F \dot f_2^4 \dot f_1^3-4 \dot f_1^2 \dot\xi^3 f_1 F \dot f_2^5-8 f_2^2 \dot f_1^3 \dot\xi^3 \dot F \dot f_2^3-16 f_2^2 \dot f_1^2 \dot\xi^3 \dot F \gamma^2 \dot f_2^4-14 f_2 \dot f_1 \dot\xi^3 \gamma^4 F \dot f_2^6\nonumber\\
&&-8 f_2^2 \dot f_1 \dot\xi^3 \dot F \gamma^4 \dot f_2^5-22 f_2 \dot f_1^2 \dot\xi^3 \gamma^2 F \dot f_2^5+f_2 \dot\xi^2 f_1 F^2 \dot f_2^6 \gamma^4-2 f_2^2 \dot f_2^3 f_1^3 \dot F^2 F \dot\xi-4 f_2^2 \dot f_1 \dot\xi^2 \dot F^2 f_1^2 \dot f_2^3-16 f_2 \dot\xi^3 \dot F f_1 \dot f_2^6 \gamma^4\nonumber\\
&&-3 f_2^5 \dot F^2 f_1 \gamma^4 F^2 \dot f_2^2-3 f_2^4 \dot f_2^2 f_1^2 \dot F^2 \gamma^2 F^2-16 f_2^3 \dot F^2 f_1 \gamma^4 \dot\xi^2 \dot f_2^4-f_2^4 \dot\xi \dot F F^2 \dot f_2 \dot f_1^3-4 f_2^2 \dot f_1^2 \dot\xi^2 F^2 \dot f_2^4 \gamma^2-4 f_2^2 \dot f_1 \dot\xi^2 F^2 \dot f_2^5 \gamma^4\nonumber\\
&&+8 f_2^4 \dot f_1^2 \dot F^2 \gamma^2 \dot\xi^2 \dot f_2^2+4 f_2^4 \dot f_1 \dot F^2 \gamma^4 \dot\xi^2 \dot f_2^3-2 \dot f_1 \dot\xi^3 f_1 \gamma^2 F \dot f_2^6-6 f_2^3 \dot f_1 \dot\xi \dot F f_1 F^2 \dot f_2^3 \gamma^2-18 f_2^4 \dot f_1 \dot F^2 f_1 \gamma^2 F \dot\xi \dot f_2^2\nonumber\\
&&-20 f_2^2 \dot f_1 \dot\xi^2 \dot F f_1 F \dot f_2^4 \gamma^2-12 \dot\xi^4 \dot f_2^5 \dot f_1^3-14 f_2^4 \dot F^2 f_1 \gamma^4 F \dot\xi \dot f_2^3+2 f_2^5 \dot f_1^2 \dot F^2 \gamma^2 F \dot\xi \dot f_2-4 f_2^4 \dot f_1^2 \dot\xi \dot F F^2 \dot f_2^2 \gamma^2\nonumber\\
&&-3 f_2 \dot f_1 \dot\xi^2 f_1 \gamma^2 \dot f_2^5 F^2-8 f_2^3 \dot F^2 f_1^2 \gamma^2 F \dot\xi \dot f_2^3-12 f_2^2 \dot\xi^2 \dot F f_1 \gamma^4 F \dot f_2^5-2 f_2^2 \dot\xi \dot F f_1^2 \gamma^2 F^2 \dot f_2^4-12 f_2^3 \dot f_1 \dot\xi^2 \dot F \gamma^4 F \dot f_2^4\nonumber\\
&&-2 f_2^3 \dot\xi \dot F f_1 \gamma^4 F^2 \dot f_2^4-20 f_2^3 \dot f_1^2 \dot\xi^2 \dot F \gamma^2 F \dot f_2^3-4 f_2^4 \dot f_1 \dot\xi \dot F F^2 \dot f_2^3 \gamma^4-32 f_2 \dot f_1 \dot\xi^3 \dot F f_1 \gamma^2 \dot f_2^5+2 f_2^5 \dot f_1 \dot F^2 \gamma^4 F \dot\xi \dot f_2^2\nonumber\\
&&-8 f_2 \dot\xi^2 \dot F f_1^2 \gamma^2 F \dot f_2^5-32 f_2^3 \dot f_1 \dot F^2 f_1 \gamma^2 \dot\xi^2 \dot f_2^3-3 f_2^5 \dot f_1 \dot F^2 f_1 \gamma^2 F^2 \dot f_2)>0\, ,
\end{eqnarray}
where $J$ has been defined in Eq.\ (\ref{eq:defJ}). It is interesting to notice that for these theories
the reduction of the degrees of freedom does not occur as the Kasner-type
background approaches the FLRW one, since $\det A$ does not either vanish 
nor diverge in the FLRW limit. In fact, for the limiting FLRW case we have
\begin{eqnarray}
\label{eq:spe1}
&&\det A
=\frac{9}4\,\frac{F+4H\dot\xi}{a^4 (1+\gamma^2)^2}{A_{22}}\,,\\
\label{eq:spe2}
&&A_{22}=\frac{12H^2(\dot F+4H^2\dot\xi)^2(F+4H\dot\xi)}{\dot F^2(\dot F+2HF+12H^2\dot\xi)^2}\, ,
\end{eqnarray}
which are both positive when $F+4H\dot\xi>0$ as well as the odd-parity mode.

A famous example is $f(R)$-gravity. The result for this particular case
is obtained by simply setting $\dot\xi=0$ in the above equations. More
explicitly, we have 
\begin{eqnarray}
 \det A&=& \frac{3F^2(\dot f_1+2\dot f_2)^2}{[F f_1 \dot f_2+2f_1f_2\dot F+Ff_2\dot f_1+2\gamma^2f_2(F\dot f_2+f_2\dot F)]^2}\, ,\\
A_{22}&=&{\frac {4F\, (  \dot f_2 ^{2} f_1 ^{2}+ \dot f_1 ^{2} f_2 ^{2}+ \dot f_2  \dot f_1 f_1 f_2 +3\, f_2 ^{2} \dot f_2 {\gamma}^{2} \dot f_1 +3\,f_1 f_2 {\gamma}^{2} \dot f_2 ^{2}+3\, f_2 ^{2}{\gamma}^{4} \dot f_2 ^{2} ) }%
{[F f_1 \dot f_2+2f_1f_2\dot F+Ff_2\dot f_1+2\gamma^2f_2(F\dot f_2+f_2\dot F)]^2}}\, .
\end{eqnarray}
In this case the sound speed is easy to evaluate, and one can verify
that all modes propagate at the speed of light. 

\subsection{Ghost crossing}

For the particular models discussed in this section, all modes become
 ghost when $F+4H\dot\xi$ crosses $0$. It might be interesting to study
 if this ghost crossing  (i.e.\ the instant of time where any of the perturbation modes becomes a ghost) can appear without the background reaching a spacetime singularity. 

An example that shows ghost crossing can be easily constructed in the
FLRW model. For definiteness we consider models specified by
$f(R,\GB)=\tilde f(R+\lambda\GB)$. In this case, the action can be
written with the aid of one auxiliary field as 
\begin{eqnarray}
S&=\frac{M_P^2}{16\pi}\int \de^4x\,\sqrt{-g}\,
\bigl[ (R+\lambda \,\GB)F-V(F) \bigr]. \label{action2}
\end{eqnarray}
Here $\xi$ in the above calculation is to be identified with $\lambda F$. Hence, the ghost
crossing occurs when $F+4\lambda H\dot F=0$. As long as the function
$V(F)$ can be chosen freely, the only non-trivial equation of motion is 
\begin{equation}
\left(1+4 \lambda H^2\right)\ddot F
-H(1-8 \lambda \dot H+4\lambda H^2 )\dot F 
+2 \dot H F=0. 
\end{equation}
As a simple example, we assume power law expansion, $a\propto t^n$. Then, $F$ is solved as 
\begin{eqnarray}
&F=&w_1\, {}_2F_1\!\left(
  - {n+1+\sqrt{n^2+10n+1}\over 4},
  -{n+1-\sqrt{n^2+10n+1}\over 4};
  -{n+1\over 2};-{t^2\over 4 n^2 \lambda}\right)\cr
 &&\quad +w_2\,  t^{n+3}\, {}_2F_1\!\left(
  {n+5-\sqrt{n^2+10n+1}\over 4},
  {n+5+\sqrt{n^2+10n+1}\over 4};
  {n+5\over 2};-{t^2\over 4 n^2 \lambda}\right), 
\end{eqnarray}
with $w_1$ and $w_2$ being constants. By choosing $w_1$ and $w_2$ appropriately, we can make models which pass through $F+4\lambda H\dot F=0$. For example, if we set $n=2$ and $w_2=0$, we have 
\begin{equation}
F + 4\lambda H \dot F
=w_1 {3 t^4 + 64 \lambda t^2 + 256 \lambda^2\over 768 \lambda^2}, 
\end{equation}
For negative $\lambda$, this crosses 0 when $t^2=-16\lambda/3$ or $-16\lambda$. Since the values of $R$ and $\GB$ are respectively given by
\begin{eqnarray} 
 &&R = \frac{6 n (2 n-1)}{t^2}, 
\cr
 &&\GB = \frac{24 (n-1) n^3}{t^4}, 
\end{eqnarray}
they are both regular at the ghost crossing points\footnote{Also other Riemann curvature invariants are finite for the power law solution $a\propto t^n$. For example, also the Kretschmann scalar is proportional to $t^{-4}$.}. Here we discussed the simple FLRW case, but the ghost crossing
without singularity generically occurs even if consider Kasner-type backgrounds\footnote{An example is given by the Lagrangian ${\cal L}=FR+\xi(F)\,\GB-\frac34F^2/\xi$, where $\xi(F)=\frac1{16}F\sqrt{(F-\lambda_2)/\lambda_1}$ which admits the solution $f_1\propto t^2$, $f_2\propto t^{-4}$, and $F=\lambda_1t^4+\lambda_2$. The ghost crossing occurs at $t^4=-\frac35\,\lambda_2/\lambda_1$, with $\lambda_1>0$ and $\lambda_2<0$, whereas the singularity is in $t=0$.}.

\section{Conclusions}

We studied the behavior of the perturbation for general modifications of gravity whose Lagrangian consists of a general function of $R$, the Ricci scalar and of $\GB$, the Gauss-Bonnet scalar, on Kasner-type backgrounds. We have shown that the existence of ghost is inevitable for generic models. However, the ghost mode can decouple from the physical spectrum in the FLRW limit. The kinetic term of this mode vanishes in this limit, which does not mean strong coupling here because the mass term does not vanish. Hence such cosmological models may have a chance to survive as an effective theory that describes only small deviation from the FLRW universe and/or the modifications of gravity tend to vanish at early times. We also found that the modified dispersion relation of the kind $\omega\propto \bm{k}^4$ first discussed in \cite{DeFelice09} is due to integrating out the ghost mode in the FLRW limit.
 
Also, we presented special cases that avoid the presence of the ghost. In this case one degree of freedom is absent from the beginning since the action satisfies a special relation, and there is no reduction of degrees of freedom in the FLRW limit. If we require this condition to be satisfied for rather generic backgrounds, the form of the Lagrangian is restricted to ${\cal L}=F(\phi)R+\xi(\phi)\GB-V(\phi)$. For such models, the ghost crossing, where the sign of the kinetic term flips for some modes, can happen. We have presented an example of ghost crossing within FLRW models without the background reaching a classical singularity.


\begin{acknowledgments}

The discussion during the workshop YITP-T-09-05 at Yukawa Institute was very useful to complete this work. We thank Prof.\ Jiro Soda for useful comments. The work of A.~D.~F.\ was supported by the Grant-in-Aid for Scientific Research Fund of the JSPS No.~09314. T.~T. is supported by the JSPS through Grants No.\ 21244033. We also acknowledge the support of the Grant-in-Aid for the Global COE Program ``The Next Generation of Physics, Spun from Universality and Emergence'' and the Grant-in-Aid for Scientific Research on Innovative Area No.\ 21111006 from the Ministry of Education, Culture, Sports, Science and Technology (MEXT) of Japan. 

\end{acknowledgments}


\begin{thebibliography}{99}

\bibitem{review}
  A.~De Felice and S.~Tsujikawa,
  arXiv:1002.4928 [gr-qc].

\bibitem{Starob}
  A.~A.~Starobinsky,
  Phys.\ Lett.\  B {\bf 91}, 99 (1980).

\bibitem{fRori}
S.~Capozziello,
Int.\ J.\ Mod.\ Phys.\  D {\bf 11}, 483 (2002);
S.~Capozziello, S.~Carloni and A.~Troisi,
Recent Res.\ Dev.\ Astron.\ Astrophys.\  {\bf 1}, 625 (2003);
S.~Capozziello, V.~F.~Cardone, S.~Carloni and A.~Troisi,
Int.\ J.\ Mod.\ Phys.\  D {\bf 12}, 1969 (2003);
S.~M.~Carroll, V.~Duvvuri, M.~Trodden and M.~S.~Turner,
Phys.\ Rev.\  D {\bf 70}, 043528 (2004).

\bibitem{fRviable}
L.~Amendola, R.~Gannouji, D.~Polarski and S.~Tsujikawa,
Phys.\ Rev.\  D {\bf 75}, 083504 (2007);
B.~Li and J.~D.~Barrow,
Phys.\ Rev.\  D {\bf 75}, 084010 (2007);
W.~Hu and I.~Sawicki,
Phys.\ Rev.\  D {\bf 76}, 064004 (2007);
A.~A.~Starobinsky,
JETP Lett.\  {\bf 86}, 157 (2007);
S.~A.~Appleby and R.~A.~Battye,
Phys.\ Lett.\  B {\bf 654}, 7 (2007); 
S.~Tsujikawa,
Phys.\ Rev.\  D {\bf 77}, 023507 (2008);
E.~V.~Linder,
arXiv:0905.2962 [astro-ph.CO];
L.~Amendola and S.~Tsujikawa,
Phys.\ Lett.\  B {\bf 660}, 125 (2008).

\bibitem{fRmatter}
S.~M.~Carroll, I.~Sawicki, A.~Silvestri and M.~Trodden,
New J.\ Phys.\  \textbf{8}, 323 (2006);
S.~Carloni, P.~K.~S.~Dunsby and A.~Troisi,
  Phys.\ Rev.\  D {\bf 77}, 024024 (2008);
Y.~S.~Song, W.~Hu and I.~Sawicki,
Phys.\ Rev.\  D {\bf 75}, 044004 (2007);
I.~Sawicki and W.~Hu,
Phys.\ Rev.\  D {\bf 75}, 127502 (2007);
R.~Bean, D.~Bernat, L.~Pogosian, A.~Silvestri and M.~Trodden,
Phys.\ Rev.\  D {\bf 75}, 064020 (2007);
T.~Faulkner, M.~Tegmark, E.~F.~Bunn and Y.~Mao,
Phys.\ Rev.\  D {\bf 76}, 063505 (2007);
L.~Pogosian and A.~Silvestri,
Phys.\ Rev.\  D {\bf 77}, 023503 (2008).

\bibitem{GaussBN}
S.~Nojiri, S.~D.~Odintsov and M.~Sasaki,
Phys.\ Rev.\  D {\bf 71}, 123509 (2005);
T.~Koivisto and D.~F.~Mota,
Phys.\ Lett.\  B {\bf 644}, 104 (2007);
T.~Koivisto and D.~F.~Mota,
Phys.\ Rev.\  D {\bf 75}, 023518 (2007);
S.~Kawai, M.~a.~Sakagami and J.~Soda,
Phys.\ Lett.\  B {\bf 437}, 284 (1998);
S.~Nojiri and S.~D.~Odintsov,
Phys.\ Lett.\  B {\bf 631}, 1 (2005);
A.~De Felice and S.~Tsujikawa,
Phys.\ Lett.\  B {\bf 675}, 1 (2009);
T.~P.~Sotiriou,
arXiv:0710.4438 [gr-qc];
K.~Uddin, J.~E.~Lidsey and R.~Tavakol,
Gen.\ Rel.\ Grav.\  {\bf 41}, 2725 (2009);
G.~Cognola, E.~Elizalde, S.~Nojiri, S.~D.~Odintsov and S.~Zerbini,
Phys.\ Rev.\  D {\bf 73}, 084007 (2006);
B.~Li, J.~D.~Barrow and D.~F.~Mota,
Phys.\ Rev.\  D {\bf 76}, 044027 (2007);
S.~Y.~Zhou, E.~J.~Copeland and P.~M.~Saffin,
JCAP {\bf 0907}, 009 (2009);
A.~De Felice and S.~Tsujikawa,
Phys.\ Rev.\  D {\bf 80}, 063516 (2009);
  A.~De Felice, D.~F.~Mota and S.~Tsujikawa,
  Phys.\ Rev.\  D {\bf 81}, 023532 (2010)
  [arXiv:0911.1811 [gr-qc]].

\bibitem{PQR}
S.~M.~Carroll, A.~De Felice, V.~Duvvuri, D.~A.~Easson, M.~Trodden and M.~S.~Turner,
Phys.\ Rev.\  D {\bf 71}, 063513 (2005);
G.~Calcagni, S.~Tsujikawa and M.~Sami,
Class.\ Quant.\ Grav.\  {\bf 22}, 3977 (2005).
  
\bibitem{DeFelice09}
A.~De Felice and T.~Suyama,
JCAP {\bf 0906}, 034 (2009);
  A.~De Felice and T.~Suyama,
  Phys.\ Rev.\  D {\bf 80}, 083523 (2009)
  [arXiv:0907.5378 [astro-ph.CO]];
  A.~De Felice, J.~M.~Gerard and T.~Suyama,
  arXiv:1005.1958 [astro-ph.CO].

\bibitem{otherFRG}
  K.~Bamba, S.~D.~Odintsov, L.~Sebastiani and S.~Zerbini,
  Eur.\ Phys.\ J.\  C {\bf 67}, 295 (2010)
  [arXiv:0911.4390 [hep-th]];
  O.~Gorbunova and L.~Sebastiani,
  arXiv:1004.1505 [gr-qc];
  E.~Elizalde, R.~Myrzakulov, V.~V.~Obukhov and D.~Saez-Gomez,
  Class.\ Quant.\ Grav.\  {\bf 27}, 095007 (2010)
  [arXiv:1001.3636 [gr-qc]].

\bibitem{ghost1}
K.~S.~Stelle,
Gen.\ Rel.\ Grav.\  {\bf 9}, 353 (1978);
N.~H.~Barth and S.~M.~Christensen,
Phys.\ Rev.\ D {\bf 28}, 1876 (1983);
N.~Boulanger, T.~Damour, L.~Gualtieri and M.~Henneaux,
  Nucl.\ Phys.\  B {\bf 597}, 127 (2001)
  [arXiv:hep-th/0007220];
A.~De Felice, M.~Hindmarsh and M.~Trodden,
JCAP {\bf 0608}, 005 (2006);
G.~Calcagni, B.~de Carlos and A.~De Felice,
Nucl.\ Phys.\ B {\bf 752}, 404 (2006).

\bibitem{Hanlon}
J.~O' Hanlon,
Phys.\ Rev.\ Lett.\  {\bf 29}, 137 (1972);
P.~Teyssandier and P.~Tourrenc,
J.\ Math.\ Phys.\  {\bf 24}, 2793 (1983);
T.~Chiba,
Phys.\ Lett.\  B {\bf 575}, 1 (2003).

\bibitem{cosmoreview}
J.~M.~Bardeen,
Phys.\ Rev.\  D {\bf 22}, 1882 (1980);
H.~Kodama and M.~Sasaki,
Prog.\ Theor.\ Phys.\ Suppl.\  {\bf 78} (1984) 1;
V.~F.~Mukhanov, H.~A.~Feldman and R.~H.~Brandenberger,
Phys.\ Rept.\  {\bf 215}, 203 (1992).


\end{thebibliography}
\end{document}